\documentclass[useAMS,usenatbib,usegraphicx]{mn2e}
\usepackage{times}

\title[Electron impact excitation of \mbox{O\,{\sc ii}} fine-structure levels]
{Electron impact excitation of \mbox{O\,{\sc ii}} fine-structure levels
\thanks{This paper is dedicated to the memory of Don Osterbrock (1924-2007) and 
Mike Seaton (1923-2007), who first calibrated the \mbox{O\,{\sc ii}} density
indicator, and did so much to advance the study of nebulae.}}
\author[R. Kisielius, P.J. Storey, G.J. Ferland, F.P. Keenan]
{R. Kisielius$^{1}$\thanks {E-mail: R.Kisielius@itpa.lt}, 
P. J. Storey$^{2}$, 
G. J. Ferland$^{3, 4}$,  
and F. P. Keenan$^{5}$
\\
$^{1}$Institute of Theoretical Physics and Astronomy,
Vilnius University, A.~Go{\v s}tauto 12,
01108 Vilnius, Lithuania\\
$^{2}$Department of Physics and Astronomy, 
University College London, Gower Street,
London WC1E 6BT, UK\\
$^{3}$Department of Physics, University of Kentucky, Lexington,
KY 40506, USA\\
$^{4}$Institute of Astronomy, University of Cambridge, Madingley
 Road, Cambridge CB3 0HA, UK\\
$^{5}$Astrophysics Research Centre,
School of Mathematics and Physics,
Queen's University Belfast,
Belfast BT7 1NN, Northern Ireland, UK}

\begin{document}

\date{\today}

\pagerange{\pageref{firstpage}--\pageref{lastpage}} \pubyear{2008}

\maketitle

\label{firstpage}

\maketitle

\begin{abstract}
Effective collision strengths for forbidden transitions among the 5 energetically
lowest fine-structure levels of \mbox{O\,{\sc ii}} are calculated in the 
Breit-Pauli approximation using the R-matrix method. Results are presented
for the electron temperature range 100 -- 100000~K. The accuracy of the 
calculations is evaluated via the use of different types of radial orbital sets
and a different configuration expansion basis for the target wavefunctions.
A detailed assessment of previous available data is given, and erroneous results
are highlighted. Our results reconfirm the validity of the original Seaton and 
Osterbrock scaling for the optical \mbox{O\,{\sc ii}} ratio, a matter of some 
recent controversy. Finally we present plasma diagnostic diagrams using the best 
collision strengths and transition probabilities.
\end{abstract}



\section{
\label{s:int}
Introduction
}

Oxygen ions in different ionization stages are abundant in a wide variety
of astrophysical objects, including planetary nebulae, stellar atmospheres,
Seyfert galaxies and the interstellar medium. In particular, emission lines 
arising from transitions among the ground state 1s$^2$2s$^2$2p$^3$ levels 
of \mbox{O\,{\sc ii}} can be utilized as a diagnostic tool for determining 
electron density ($n_{\mathrm e}$). 
\citet{so1957} suggested the use of the emission doublet-line ratio  
$I(3729$\AA$)/ I(3726 $\AA$)$ of \mbox{O\,{\sc ii}} arising from nebular
transitions from the ground-state levels $^2D_{5/2}$ and
$^2D_{3/2}$ to the lowest level $^4S_{3/2}$  as a density indicator.
Their work combined Seaton's newly developed collision theories with 
Osterbrock's access to modern instrumentation to usher in a new era of precision
nebular astrophysics (\citealt{deo2000}; \citealt{agn3}).
Osterbrock's observations showed that, in the low-density limit, the observed 
\mbox{O\,{\sc ii}} line ratio was equal to the ratio of statistical 
weights of the upper levels, as expected from Seaton's theories.
The \mbox{O\,{\sc ii}} ratio was the main density indicator for nebulae until
improvement in detector technology made the red \mbox{S\,{\sc ii}} lines 
accessible. When both nebular and auroral \mbox{O\,{\sc ii}} transitions 
(at 7720 and 7730\AA) are considered, both $n_{\mathrm e}$ and the electron 
temperature $T_{\mathrm e}$  of the plasma may simultaneously be found, 
as shown by for example \citet{fpk1999}.

To calculate reliable line ratios, one must employ highly accurate atomic
data, especially for electron impact excitation rates and
transition probabilities for the forbidden lines. Until the last decade,
the most reliable excitation rates for transitions among the
1s$^2$2s$^2$2p$^3$ levels of \mbox{O\,{\sc ii}} have been those of 
\citet{pradhan76}, obtained by employing the R-matrix method with inclusion
of the five energetically lowest $LS$ states, 1s$^2$2s$^2$2p$^3$ $^4S$, $^2D$,
$^2P$, and 1s$^2$2s2p$^4$ $^4P$, $^2D$. Although the calculation was performed
in the non-relativistic approach, the data for the excitation of the fine-structure
levels 1s$^2$2s$^2$2p$^3$ $^2D_{3/2}$, $^2D_{5/2}$ from the ground state
$^4S_{3/2}$ were customarily obtained by splitting the non-relativistic values
of the excitation rates $\Upsilon$  proportionally to the statistical weights
of the final levels, the scaling originally suggested by \citet{so1957}.

However, \citet{BMM98} have recalculated excitation rates for \mbox{O\,{\sc ii}} 
using the R-matrix method within the Breit-Pauli approximation, where
the 11 fine-structure levels were included explicitly into a close-coupling
formulation of the scattering problem. Their data are significantly different
from those of \citet{pradhan76}, and the differences were attributed to
the larger number of states included and a better resolution of the resonance
structure in the calculation of McLaughlin \& Bell.
Subsequently, \citet{fpk1999} used these newly calculated electron
impact excitation rates in their model to calculate the emission-line ratio
diagrams for lines of \mbox{O\,{\sc ii}} for a range of $T_{\mathrm e}$
and $n_{\mathrm e}$ appropriate to gaseous nebulae.

More recently, \citet{copetti} compared density estimates for planetary nebulae 
based on different density-indicator lines of \mbox{O\,{\sc ii}},
\mbox{S\,{\sc ii}}, \mbox{Cl\,{\sc iii}}, \mbox{Ar\,{\sc iv}}, \mbox{C\,{\sc iii}} 
and \mbox{N\,{\sc i}}. They found systematic deviations for values of $n_{\mathrm e}$ 
derived from the \mbox{O\,{\sc ii}} lines, and attributed these to errors 
in the atomic data, particularly the collision strengths used by \citet{fpk1999}. 
Furthermore, \cite{wangliu} have considered four density indicators, including
the [\mbox{O\,{\sc ii}}] $\lambda 3729/ \lambda 3726$ doublet ratio, for a large 
sample of more than 100 planetary nebulae, and concluded that the calculations
of collision strengths by \cite{BMM98} are inconsistent with the observations.

Very recently, \citet{mm2006} have investigated relativistic and correlation 
effects in electron-impact excitation of \mbox{O\,{\sc ii}} using 
the Breit-Pauli R-matrix method. They concluded that the fine-structure 
collision strengths are not affected by relativistic effects and do not 
significantly depart from the values obtained from a $LS \rightarrow LSJ$ 
transformation. \citet{ap2005} discussed the astrophysical implications 
of these new atomic data and have derived the \mbox{O\,{\sc ii}} line ratios 
$I(3729)/I(3726)$. Their results confirmed analyses of \citet{copetti} 
and \cite{wangliu}. 
Furthermore, \citet{tayal2006} and \citet{tayal2007} have reported similar 
calculation for \mbox{O\,{\sc ii}}, employing the B-spline R-matrix method with
non-orthogonal sets of radial functions and the inclusion of 47 fine-structure 
levels. This author also performed a Breit-Pauli R-matrix calculation with 
orthogonal radial functions involving 62 fine-structure levels in the 
close-coupling expansion, as an independent check on cross sections for the 
forbidden and allowed transitions in \mbox{O\,{\sc ii}}.
 
In our work we study electron-impact excitation of forbidden lines in
\mbox{O\,{\sc ii}} using the R-matrix approach in the Breit-Pauli framework. We
attempt to establish if relativistic effects and a sufficient resolution
of the resonance structure in the collision strengths can cause the significant
departure from the statistical-weights ratio for the Maxwellian-averaged
effective collision strengths $\Upsilon$, as was claimed in \citet{BMM98}. 
Two different sets of configuration basis are employed to describe the target 
states, in order to evaluate the influence of the number of states
included in the scattering problem on the collision strength parameters.
Furthermore, we use two different types of radial orbitals, namely those obtained
using Thomas-Fermi-Dirac model potential and Slater-type orbitals, in our
scattering calculation.
We present a comparison of our calculated energy levels, multiplet oscillator
strengths and effective collision strengths obtained using different
configuration sets and different radial orbitals with both available experimental
data and the theoretical results of other authors.

\section{
\label{s:adc}
Atomic data calculation
}

In the present work we determine electron-impact collision strengths for
the electric-dipole forbidden transitions among the five lowest levels
of \mbox{O\,{\sc ii}}. All possible excitation processes among the fine-structure
levels $^4S_{3/2}^{\mathrm o}$, $^2D_{5/2}^{\mathrm o}$,
$^2D_{3/2}^{\mathrm o}$, $^2P_{3/2}^{\mathrm o}$, $^2P_{1/2}^{\mathrm o}$
of the ground configuration 1s$^2$2s$^2$2p$^3$ are examined
using R-matrix close-coupling codes.
Collision strengths ($\Omega$) are calculated using a very fine energy mesh
for the impact electron energies from the first excitation threshold
to the highest threshold, and a coarse energy mesh in the region above all
thresholds. These data are thermally averaged for effective
collision strengths ($\Upsilon$) to be determined in the temperature range
100 -- 100\,000 K.

\subsection{
\label{s:target}
The scattering target
}

\begin{table*}
 \centering
 \begin{flushleft}
  \caption{Fine-structure energy levels, their indices N, experimental
           and calculated energies (Ry) for \mbox{O\,{\sc ii}} relative
          to the ground level 2s$^2$2p$^3$ $^4$S$_{3/2}^{\mathrm o}$}.
  \label{tab:target}
  \begin{tabular}{rlllrlrlrll}
  \hline
  \multicolumn{1}{c}{N}&
  \multicolumn{1}{l}{Level}&
  \multicolumn{1}{l}{$E_{\mathrm{exp}}$}&
  \multicolumn{1}{l}{$E_{\mathrm{TFD}}$}&
  \multicolumn{1}{l}{$\Delta E_{\mathrm{TFD}}$}&
  \multicolumn{1}{l}{$E_{\mathrm{TFD1}}$}&
  \multicolumn{1}{l}{$\Delta E_{\mathrm{TFD1}}$}&
  \multicolumn{1}{l}{$E_{\mathrm{STO1}}$}&
  \multicolumn{1}{l}{$\Delta E_{\mathrm{STO1}}$}&
  \multicolumn{1}{l}{$E_{\mathrm{MB98}}$}&
  \multicolumn{1}{l}{$E_{\mathrm{T07}}$}\\
  \hline
 1& 2s$^2$2p$^3$ $^4$S$_{3/2}^{\mathrm o}$&                    
0.00000&  0.00000&  0.00000 &  0.00000&  0.00000&  0.00000&  0.00000& 0.00000& 0.00000\\
 2& 2s$^2$2p$^3$ $^2$D$_{5/2}^{\mathrm o}$&                    
0.24432&  0.24596& -0.00164 &  0.25930& -0.01498&  0.26519& -0.02087& 0.24789& 0.25349\\
 3& 2s$^2$2p$^3$ $^2$D$_{3/2}^{\mathrm o}$&                    
0.24450&  0.24583& -0.00133 &  0.25916& -0.01466&  0.26507& -0.02057& 0.24791& 0.25369\\
 4& 2s$^2$2p$^3$ $^2$P$_{3/2}^{\mathrm o}$&                    
0.36877&  0.37911& -0.01034 &  0.37839& -0.00962&  0.38667& -0.01790& 0.37413& 0.38389\\
 5& 2s$^2$2p$^3$ $^2$P$_{1/2}^{\mathrm o}$&                    
0.36879&  0.37899& -0.01020 &  0.37822& -0.00943&  0.38653& -0.01774& 0.37410& 0.38387\\
 6& 2s2p$^4$ $^4$$P_{5/2}^{\mathrm e}$&                        
1.09204&  1.06500&  0.02704 &  1.05995&  0.03209&  1.06824&  0.02380& 1.12351& 1.10092\\
 7& 2s2p$^4$ $^4$$P_{3/2}^{\mathrm e}$&                        
1.09353&  1.06693&  0.02660 &  1.06188&  0.03165&  1.07010&  0.02343& 1.12480& 1.10230\\
 8& 2s2p$^4$ $^4$$P_{1/2}^{\mathrm e}$&                        
1.09428&  1.06808&  0.02620 &  1.06303&  0.03125&  1.07121&  0.02307& 1.12456& 1.10313\\
 9& 2s2p$^4$ $^2$$D_{5/2}^{\mathrm e}$&                        
1.51260&  1.50976&  0.00284 &  1.51245&  0.00015&  1.52500& -0.01240& 1.54667& 1.54369\\
10& 2s2p$^4$ $^2$$D_{3/2}^{\mathrm e}$&                        
1.51267&  1.50959&  0.00308 &  1.51263&  0.00004&  1.52483& -0.01216& 1.54660& 1.54375\\
11& 2s$^2$2p$^2$3s $^4$P$_{1/2}^{\mathrm e}$&                  
1.68799&  1.69757& -0.00958 &  1.68990& -0.00191&  1.69592& -0.00793& 1.67762& 1.69192\\
12& 2s$^2$2p$^2$3s $^4$P$_{3/2}^{\mathrm e}$&                  
1.68895&  1.69910& -0.01015 &  1.69143& -0.00248&  1.69736& -0.00841& 1.67859& 1.69277\\
13& 2s$^2$2p$^2$3s $^4$P$_{5/2}^{\mathrm e}$&                  
1.69039&  1.70163& -0.01124 &  1.69398& -0.00359&  1.69976& -0.00937& 1.68021& 1.69418\\
14& 2s$^2$2p$^2$3s $^2$P$_{1/2}^{\mathrm e}$&                  
1.72128&  1.74231& -0.02103 &  1.73882& -0.01754&  1.72950& -0.00822& 1.72595& 1.72653\\
15& 2s$^2$2p$^2$3s $^2$P$_{3/2}^{\mathrm e}$&                  
1.72128&  1.74501& -0.02373 &  1.74163& -0.02035&  1.73219& -0.01091& 1.72778& 1.72811\\
16& 2s2p$^4$ $^2$S$_{1/2}^{\mathrm e}$&                        
1.78345&  1.80089& -0.01744 &  1.79652& -0.01307&  1.81621& -0.03276& 1.83618& 1.79711\\
17& 2s$^2$2p$^2$3s$^{\prime}$ $^2$D$_{5/2}^{\mathrm e}$&       
1.88606&  1.92459& -0.03853 &  1.91654& -0.03048&  1.92018& -0.03412& 1.90172& 1.90193\\
18& 2s$^2$2p$^2$3s$^{\prime}$ $^2$D$_{3/2}^{\mathrm e}$&       
1.88607&  1.92460& -0.03853 &  1.91655& -0.03048&  1.92018& -0.03411& 1.90173& 1.90194\\
19& 2s2p$^4$ $^2$P$_{3/2}^{\mathrm e}$&                        
1.93730&  1.98880& -0.05150 &  1.98950& -0.05220&  1.99860& -0.06130& 2.08127& 1.96267\\
20& 2s2p$^4$ $^2$P$_{1/2}^{\mathrm e}$&                        
1.93883&  1.99094& -0.05211 &  1.99172& -0.05289&  2.00082& -0.06199& 2.08283& 1.96424\\
21& 2s$^2$2p$^2$3s$^{\prime\prime}$ $^2$S$_{1/2}^{\mathrm e}$& 
2.10147&  2.18357& -0.08210 &  2.18317& -0.08170&  2.18639& -0.08492& 2.17157& 2.12495\\
\hline
\end{tabular}
\end{flushleft}
\end{table*}

In the present work we use two different sets of configuration basis
describing the O$^+$ target states.
We include odd configurations 2s$^2$2p$^3$, 2p$^5$, 2s$^2$2p$^2$$\overline{3}$p,
2s$^2$2p$^2$$\overline{4}$f, 2s2p$^3$$\overline{3}$d, 2p$^4$$\overline{3}$p,
2s$^2$2p$\overline{3}$p$^2$, 2s$^2$2p$\overline{3}$d$^2$, 2p$^3$$\overline{3}$d$^2$
and even configurations 2s2p$^4$, 2s$^2$2p$^2$3s, 2s$^2$2p$^2$$\overline{3}$d,
2s2p$^3$$\overline{3}$p, 2s2p$^3$$\overline{4}$f, 2p$^4$3s, 2p$^4$$\overline{3}$d,
2s$^2$2p3s$\overline{3}$p, 2s2p$^2$3s$^2$, 2s2p$^2$$\overline{3}$p$^2$,
2s2p$^2$$\overline{3}$d$^2$, 2s2p$^2$3s$\overline{3}$d, 2p$^3$3s$\overline{3}$p
for the basis wavefunction configuration-interaction (CI) expansion in our
larger calculation, denoted later as {\sl TFD}.
For the scattering problem only the lowest 11 $LS$ terms are included,
which give rise to 21 fine-structure levels.
The target wave functions are calculated using the general purpose
atomic structure code {\sc superstructure} (\citealt{ss74},
\citealt{ns78}). The one-electron radial functions
were calculated in adjustable Thomas-Fermi-Dirac model potentials,
with the potential scaling parameters $\lambda_{nl}$ determined
by minimizing the sum of the energies of the 11 target states in $LS$-coupling.
In our case, we obtained values for the scaling parameters of:
$\lambda_{1s}$ = 1.465,
$\lambda_{2s}$ = 1.175,
$\lambda_{2p}$ = 1.129,
$\lambda_{3s}$ = 1.326,
$\lambda_{{\overline3}p}$ = -0.785,
$\lambda_{{\overline3}d}$ = -1.044,
$\lambda_{{\overline4}f}$ = -1.646,
with the negative values having the significance detailed by
\citet{ns78}.

In Table~\ref{tab:target} we compare experimental target state energies
$E_{\mathrm{exp}}$ \citep{wen90} with our values
$E_{\mathrm{TFD}}$ obtained using the above-described wavefunctions
for the O$^+$ target. Energies are presented relative to the ground level
1s$^2$2s$^2$2p$^3$ $^4$S$_{3/2}^{\mathrm o}$. In addition, we list the energy
differencies $\Delta E_{\mathrm{TFD}} = E_{\mathrm{exp}}- E_{\mathrm{TFD}}$, 
which were used in the scattering calculation to adjust the theoretical levels 
so that they match the experimental ones, ensuring a more accurate resonance 
positioning. There is clearly very good agreement between the calculated and 
observed energy levels. In most cases the difference is 1--2\% or less, 
and even for the highest level 
1s$^2$2s$^2$2p$^2$3s$^{\prime\prime}$ $^2$S$_{1/2}^{\mathrm e}$
the discrepancy is only 3.9\%.

A smaller set consisting of the configurations 2s$^2$2p$^3$, 2p$^5$,
2s$^2$2p$^2$$\overline{3}$p, 2s2p$^3$$\overline{3}$d, 2s$^2$2p$\overline{3}$p$^2$,
2s$^2$2p$\overline{3}$d$^2$ for odd symmetries and configurations 2s2p$^4$,
2s$^2$2p$^2$3s, 2s$^2$2p$^2$$\overline{3}$d, 2s2p$^3$$\overline{3}$p, 2p$^4$3s,
2p$^4$$\overline{3}$d, 2s$^2$2p3s$\overline{3}$p, 2s2p$^2$3s$^2$, 2s2p$^2$$\overline{3}$p$^2$,
2s2p$^2$$\overline{3}$d$^2$, 2s2p$^2$3s$\overline{3}$d for even symmetries
is introduced to replicate the calculation of \citet{BMM98},
and to check the convergence of our calculations. This set differs
from the larger one mainly by the omission of configurations containing the
$\overline{4}f$ orbital. We use two different sets of one-electron radial
orbitals for this set of configurations. In the first, denoted
as {\sl TFD1}, we utilize the same radial functions as in the {\sl TFD} calculation, while
in the second set (denoted {\sl STO1}) we use Slater-type radial orbitals
obtained employing the {\sc civ3} code of \citet{civ3}.
Their parameters were determined by \citet{bell89} for a photoionization 
calculation, and were used by \citet{BMM98} in the electron-impact 
excitation calculation of O$^+$.

Similarly to the previous set, only the lowest 11 $LS$ terms yielding
21 fine-structure levels are included in the scattering calculation.
Target level energies obtained with this set are denoted as
$E_{\mathrm{TFD1}}$ and $E_{\mathrm{STO1}}$, and are presented in
Table~\ref{tab:target} together with the energy differences
$\Delta E_{\mathrm{TFD1}} = E_{\mathrm{exp}}- E_{\mathrm{TFD1}}$ and
$\Delta E_{\mathrm{STO1}} = E_{\mathrm{exp}}- E_{\mathrm{STO1}}$.

The energy levels calculated by \cite{BMM98} are presented in the column 
$E_{\mathrm{MB98}}$ of Table~\ref{tab:target}.
One can see some differences between the level energies
$E_{\mathrm{STO1}}$ and $E_{\mathrm{MB98}}$, which can be explained
by the different number of configuration state functions (CSFs)
used in these calculations. We employ a complete set of CSFs arising
from the configurations included in the CI wavefunction expansion,
whereas \cite{BMM98} include a restricted number of
CSFs in the wavefunction representation of the O$^+$ states
(for more details see \citealt{bell89}). Finally, in the last column of
Table~\ref{tab:target} we present the energy levels from \citet{tayal2007},
denoted ($E_{T07}$), calculated using non-orthogonal B-spline radial functions.
We note that our calculated energy levels for the ground 2s$^2$2p$^3$
configuration of \mbox{O\,{\sc ii}} are closer to the experimental values
comparing to the data of \citet{tayal2007} but this is not true for the levels 
of excited configurations. Since we are dealing with the transitions within
the ground configuration, these deviations does not play a significant role 
on the accuracy of calculated collision strengths.

One of the ways to estimate the accuracy of chosen wavefunctions is to
compare the length and the velocity forms of the multiplet oscillator
strengths for electric dipole (E1) transitions calculated in $LS$-coupling.
In Table~\ref{tab:gf} we compare our data with the results from \citet{tayal2007} 
(T07) and with those from the more elaborate calculation of \citet{bell94} 
which employs the {\sc civ3} code. Since the latter $gf$-values were obtained 
in the Breit-Pauli approximation for all lines in the multiplet, we have 
averaged them in order to obtain oscillator strengths for multiplets.
Our calculation was performed using the {\sc superstructure} code of
\citet{ss74} in non-relativistic $LS$-coupling.
One can see that, in general, there is reasonable agreement between
our $gf_{\mathrm L}$ and $gf_{\mathrm V}$ values.
The 2p$^3$ $^4$S$^{\rm o}$ -- 2s2p$^4$ $^4$P resonance transition
shows a greater discrepancy between length anf velocity forms but here
the length result is in good agreement with the result of \citet{bell94}. 
Indeed, there is generally good agreement between our results and those 
of \cite{bell94} in the length formulation.

The only considerable discrepancy exists for
the 2p$^3$ $^2$P$^{\mathrm o}$ -- 2s2p$^4$ $^2$P multiplet, where the data differs
by a factor of 2. There is a similar discrepancy between the two sets of data  when
we calculate $gf$-values for the fine-structure lines within
the Breit-Pauli approximation. Nevertheless, we conclude that the CI wavefunctions
employed in our {\sl TFD} set of calculation are of high accuracy.

\subsection{
\label{s:excitation}
The scattering calculation
}

We apply the R-matrix method within the Breit-Pauli (BP) approximation as 
described in \citet{rm75}, \citet{rm80} and \citet{seaton87}, and implemented 
by \citet{rm87}, \citet{rm95} to compute collision strengths $\Omega$ for 
electron-impact on the O$^+$ ion.
In this approach, a non-relativistic Hamiltonian is extended to
explicitly include one-electron relativistic terms from the Breit-Pauli
Hamiltonian, namely the spin-orbit interaction term, the mass-correction term
and the one-electron Darwin term. We use an R-matrix boundary radius of
15.0 a.u. to contain the most diffuse target orbital 3s. The radial orbitals
${\overline3}$p, ${\overline3}$d and ${\overline4}$f describing pseudo-states 
are orthogonalised to bound orbitals using the Schmidt procedure.
Expansion of each scattered electron partial wave is over the basis
of 25 continuum wavefunctions within the R-matrix boundary, and the
Buttle corrections are added to compensate for the truncation to the
finite number of terms in the R-matrix expansion. This allows us
to compute accurate collision data for electron energies up to
15 Ry. The partial wave expansion for the (N+1)-electron system extends
to a maximum total angular momentum  $L = 12$ and includes
singlet, triplet and quintet $LS$ symmetries for both even and
odd parities. Subsequently, the Hamiltonian matrices and the long-range
potential coefficients obtained in a $LS$-coupling are transformed
by means of a unitary transformation to a pair-coupling scheme.
The intermediate coupling Hamiltonian matrices are
then calculated for the even and odd parities up to a total angular
momentum $J = 10$, with the theoretical target level energies adjusted
by $\Delta E_{\mathrm{TFD}}$ (see Table~\ref{tab:target}) to ensure
they match the observed values. We perform the full exchange R-matrix
outer region calculation for values of $J = 0-10$, and
top-up these data for non-dipole allowed transitions assuming that
the collision strengths form a geometric progression in $J$ for
$J > 10$. 
In practice the collision strengths for the transitions between the
fine-structure levels are already well converged by $J=10$.  

For example, considering the transition $^4$S$_{3/2}^{\mathrm o}$ $-$
$^2$D$_{5/2}^{\mathrm o}$ $\Omega$ at the $E=2.1$~Ry, the partial
waves with $J=0-5$ contribute $99.8\%$ of the total collision
strength. Similar behaviour is seen for the other trasitions and
convergence is even faster at lower energies.

\begin{table}
 \centering
 \begin{flushleft}
  \caption{
Comparison of weighted multiplet oscillator strengths in the length
          ($gf_{\mathrm L}$) and velocity ($gf_{\mathrm V}$) forms obtained
           in our calculation using the {\sc superstructure} code (SS)
           with the data from \citet{bell94} (CIV3) and \citet{tayal2007}
	   (T07)
}
  \label{tab:gf}
  \begin{tabular}{llllll}
  \hline
  &
  \multicolumn{2}{c}{\sc{ss}}&
  \multicolumn{2}{c}{\sc{civ3}}&
  \multicolumn{1}{c}{\sc{T07}}\\
  \cline{2-6}
  \multicolumn{1}{c}{Multiplet}&
  \multicolumn{1}{l}{$gf_{\mathrm L}$}&
  \multicolumn{1}{l}{$gf_{\mathrm V}$}&
  \multicolumn{1}{l}{$gf_{\mathrm L}$}&
  \multicolumn{1}{l}{$gf_{\mathrm V}$}&
  \multicolumn{1}{l}{$gf_{\mathrm L}$}\\
  \hline
2p$^3$ $^4$S$^{\mathrm o}$ -- 2s2p$^4$ $^4$P & 1.068& 1.695& 1.100& 1.240& 1.200\\
2p$^3$ $^4$S$^{\mathrm o}$ -- 2p$^2$3s $^4$P & 0.508& 0.450& 0.508& 0.500& 0.448\\
2p$^3$ $^2$D$^{\mathrm o}$ -- 2s2p$^4$ $^2$D & 1.726& 2.251& 1.540& 1.710& 1.820\\
2p$^3$ $^2$D$^{\mathrm o}$ -- 2p$^2$3s $^2$P & 1.443& 1.564& 1.200& 1.250& 1.046\\
2p$^3$ $^2$D$^{\mathrm o}$ -- 2p$^2$3s $^2$D & 0.542& 0.487& 0.510& 0.510& 0.404\\
2p$^3$ $^2$D$^{\mathrm o}$ -- 2s2p$^4$ $^2$P & 1.791& 1.878& 1.500& 1.600& 1.526\\
2p$^3$ $^2$P$^{\mathrm o}$ -- 2s2p$^4$ $^2$D & 0.235& 0.359& 0.186& 0.222& 0.244\\
2p$^3$ $^2$P$^{\mathrm o}$ -- 2p$^2$3s $^2$P & 0.241& 0.282& 0.234& 0.216& 0.240\\
2p$^3$ $^2$P$^{\mathrm o}$ -- 2s2p$^4$ $^2$S & 0.687& 0.811& 0.678& 0.750& 0.528\\
2p$^3$ $^2$P$^{\mathrm o}$ -- 2p$^2$3s $^2$D & 0.338& 0.356& 0.270& 0.264& 0.270\\
2p$^3$ $^2$P$^{\mathrm o}$ -- 2s2p$^4$ $^2$P & 1.173& 1.322& 0.516& 0.600& 0.438\\
2p$^3$ $^2$P$^{\mathrm o}$ -- 2p$^2$3s $^2$S & 0.062& 0.046& 0.054& 0.060& 0.054\\
\hline	
\end{tabular}
\end{flushleft}
\end{table}

\section{
\label{s:results}
Results and discussion
}

We present collision strengths $\Omega$ and thermally-averaged
effective collision strengths $\Upsilon$ for the optically-forbidden
transitions among
the fine-structure levels of the ground configuration
of the O$^+$ ion. The total collision strength $\Omega_{ij}$
is symmetric in $i$ and $j$ and is given by
\begin{equation} \label{eq:omega}
\Omega_{ij} = \sum_{J\pi}\Omega_{ij}^{J\pi},
\end{equation}
where $\Omega_{ij}^{J\pi}$ is a partial collision strength for a
transition from an initial target state denoted by $\alpha_iJ_i$
to a final target state $\alpha_jJ_j$, $\alpha_i$ and $\alpha_j$
being the additional quantum numbers necessary for definition of
the target states and sum runs over all partial waves $J\pi$.

The total cross section for the transition from $i$ to $j$ can be
calculated from $\Omega_{ij}$ by the relation
\begin{equation} \label{eq:xs}
\sigma_{ij} = \frac{\pi a_0^2}{(2J_i + 1)k_i^2}\Omega_{ij}
\end{equation}
where $k_i^2$ is the scattering electron energy (in Ry) relative
to the state $i$. Note that $\sigma_{ij}$ is not symmetrical
in relation to $i$ and $j$.

Assuming that the scattering electrons have a Maxwellian velocity
distribution, we can compute the dimensionless thermally averaged or
effective collision strength $\Upsilon_{ij}$ for a transition $i
\rightarrow j$ which relates to $\Omega_{ij}(E_j)$:
\begin{equation} \label{eq:ups}
\Upsilon_{ij}(T) = \int_{0}^{\infty}\Omega_{ij}(E_j)%
                   \mathrm{e}^{-E_j/kT}d(E_j/kT)
\end{equation}
where $E_j$ is the kinetic energy of the outgoing electron, $T$ is the
electron temperature, $k$ is Boltzmann's constant, and
$\Upsilon_{ij} = \Upsilon_{ji}$. Having determined
$\Upsilon_{ij}(T)$, one can subsequently obtain the excitation rate
coefficient $q_{ij}$ (in $cm^3s^{-1}$) which is usually used in
astrophysical and plasma applications:
\begin{equation} \label{eq:ex}
q_{ij} = 8.63 \times 10^{-6}(2J_i+1)^{-1}T^{-1/2} \Upsilon_{ij}(T)%
         \mathrm{e}^{-\Delta E_{ij}/kT}.
\end{equation}
The corresponding de-excitation rate coefficient $q_{ji}$ is
\begin{equation} \label{eq:dex}
q_{ji} = 8.63 \times 10^{-6}(2J_j+1)^{-1}T^{-1/2} \Upsilon_{ij}(T)
\end{equation}
where $\Delta E_{ij}$ is the energy difference between the initial
state $i$ and the final state $j$, and $T$ is the electron temperature
in K.

\subsection{
\label{s:omega}
Collision strengths
}

\begin{figure*}
\includegraphics[width=14.4cm]{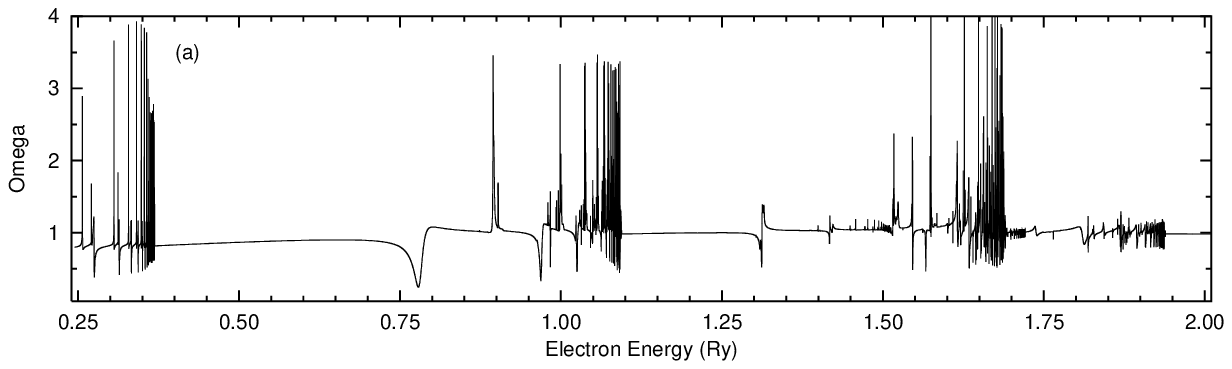}
\includegraphics[width=14.4cm]{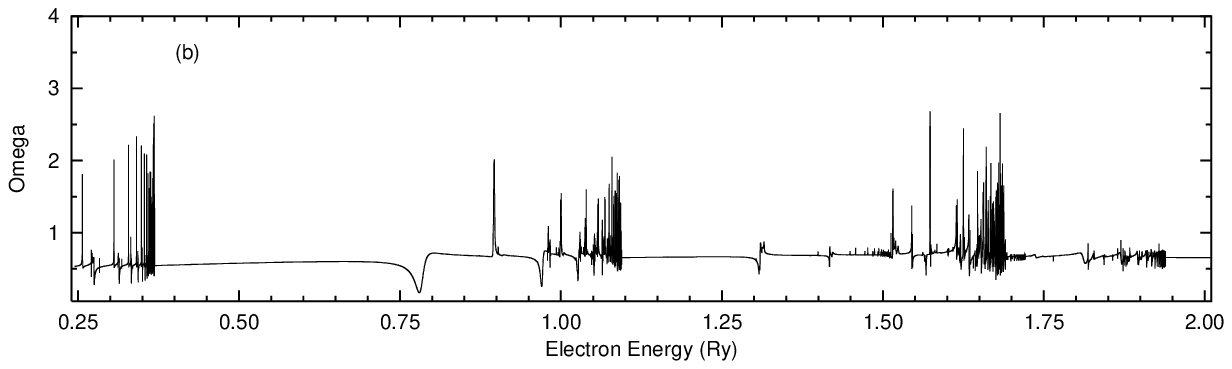}
\includegraphics[width=14.4cm]{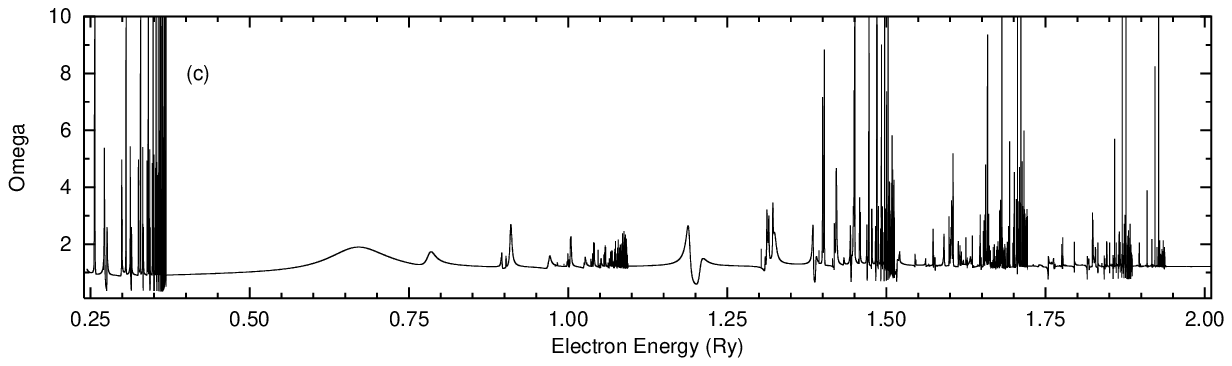}
\includegraphics[width=14.4cm]{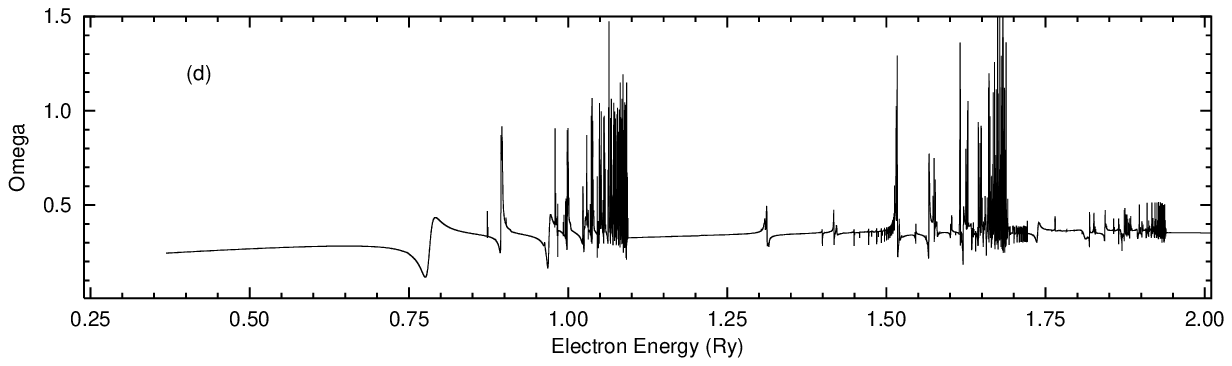}
\includegraphics[width=14.4cm]{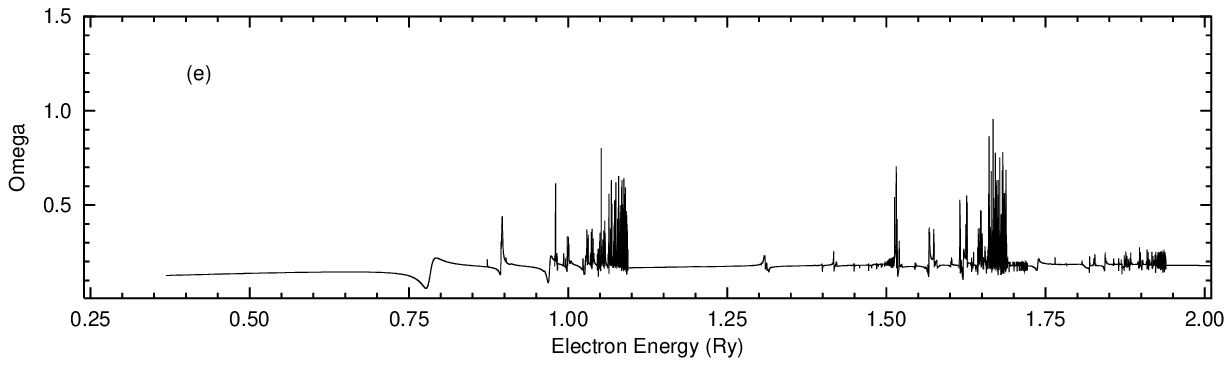}
\caption{Collision strengths  for transitions
among the fine-structure levels of the ground configuration
1s$^2$2s$^2$2p$^3$ of \mbox{O\,{\sc ii}};
(a): $^4$S$_{3/2}^{\mathrm o}$ -- $^2$D$_{5/2}^{\mathrm o}$,
(b): $^4$S$_{3/2}^{\mathrm o}$ -- $^2$D$_{3/2}^{\mathrm o}$,
(c): $^2$D$_{5/2}^{\mathrm o}$ -- $^2$D$_{3/2}^{\mathrm o}$,
(d): $^4$S$_{3/2}^{\mathrm o}$ -- $^2$P$_{3/2}^{\mathrm o}$,
(e): $^4$S$_{3/2}^{\mathrm o}$ -- $^2$P$_{1/2}^{\mathrm o}$.
Electron energies are in Rydbergs relative to the
lowest level $^4$S$_{3/2}^{\mathrm o}$.
\label{fig:omega12345}}
\end{figure*}

In the present work, the electron scattering calculation in the external
region  using a very fine energy mesh of $\Delta E = 2.5 \times 10^{-6}$~Ry
is performed for electron energies between the first excitation threshold
at 0.24432~Ry and just above the highest threshold included in the target
at 2.10147~Ry. This fine energy mesh allows us to accurately
delineate the resonance structure of the collision strengths.
For electron energies above all excitation thresholds up to 15~Ry,
a coarse energy mesh of 0.5~Ry is applied.

In Fig.~\ref{fig:omega12345} we present collision strengths
$\Omega$ for  transitions from the ground level
2s$^2$2p$^3$~$^4$S$_{3/2}^{\mathrm o}$ to the excited levels
$^2$D$_{5/2}^{\mathrm o}$, $^2$D$_{3/2}^{\mathrm o}$
$^2$P$_{3/2}^{\mathrm o}$ and $^2$P$_{1/2}^{\mathrm o}$, plus
from the
$^2$D$_{5/2}^{\mathrm o}$ level to $^2$D$_{3/2}^{\mathrm o}$
of the same configuration. One can clearly see from
Fig.~\ref{fig:omega12345} that the ratio of collisions strengths
$\Omega$($^4$S$_{3/2}^{\mathrm o}$ -- $^2$D$_{5/2}^{\mathrm o}$) /
$\Omega$($^4$S$_{3/2}^{\mathrm o}$ -- $^2$D$_{3/2}^{\mathrm o}$)
is equal to approximately $1.5$ throughout the
energy
region. This corresponds to the ratio of statistical weights
of the upper levels. A similar situation is found for the
ratio $\Omega$($^4$S$_{3/2}^{\mathrm o}$ -- $^2$P$_{3/2}^{\mathrm o}$) /
$\Omega$($^4$S$_{3/2}^{\mathrm o}$ -- $^2$P$_{1/2}^{\mathrm o}$),
which is very close to $2.0$, corresponding to the ratio of
statistical weights of the levels of the excited-state term
$^2$P$^{\mathrm o}$.

\subsection{
\label{s:upsilon}
Effective collision strengths
}

Collision strengths $\Omega$ were initially computed for
incident electron energies up to 15~Ry.  However, our target basis
contains the correlation radial orbitals ${\overline3}p$,
${\overline3}d$ and ${\overline4}f$, which we find to give rise to
pseudo-resonances at energies above $2.9$~Ry. Therefore we can not
exploit the complete energy range provided by our continuum radial
orbitals for the calculation of $\Upsilon$. Consequently, for the
purpose of computing effective collision strengths, we choose to
truncate the collision strengths $\Omega$ at a cut-off energy
$E_c=2.5$~Ry. We also computed the effective collision strengths
assuming that $\Omega$ is constant for $E > E_c$ and equal to the
value at that energy. At the upper limit of temperature for the
tabulated $\Upsilon$ values the results of the two approximations
differ by no more than $1\%$ for any transition. This gives us
confidence that the collision strengths above $2.5$~Ry can be
neglected safely.
 
In Table~\ref{tab:upstfd} we
present the calculated effective collision strengths $\Upsilon$ for
transitions among fine-structure levels of the ground configuration
1s$^2$2s$^2$2p$^3$ of \mbox{O\,{\sc ii}}. The level indices denoting a
transition correspond to the values of $N$ in
Table~\ref{tab:target}. Effective collision strengths are presented
for the temperature range $T = 100 - 100\,000$~K for all transitions
involving the five lowest levels of the ground configuration.

\begin{table}
\begin{flushleft}
\caption{
Effective collision strengths $\Upsilon$
for transition among fine-structure levels of the ground
configuration 1s$^2$2s$^2$2p$^3$ in \mbox{O\,{\sc ii}}.}
  \label{tab:upstfd}
  \begin{tabular}{rlllll}
  \hline
  \multicolumn{1}{r}{T (K)}&
  \multicolumn{1}{c}{$1 - 2$}&
  \multicolumn{1}{c}{$1 - 3$}&
  \multicolumn{1}{c}{$1 - 4$}&
  \multicolumn{1}{c}{$1 - 5$}&
  \multicolumn{1}{c}{$2 - 3$}\\
\hline
     100 &  0.796&  0.531&  0.244&  0.126&  1.095\\
     150 &  0.797&  0.533&  0.245&  0.126&  1.086\\
     200 &  0.798&  0.533&  0.245&  0.126&  1.078\\
     300 &  0.801&  0.535&  0.245&  0.126&  1.072\\
     500 &  0.808&  0.540&  0.245&  0.127&  1.097\\
     750 &  0.817&  0.546&  0.246&  0.127&  1.151\\
    1000 &  0.823&  0.550&  0.246&  0.127&  1.194\\
    1500 &  0.830&  0.554&  0.247&  0.127&  1.239\\
    2000 &  0.832&  0.555&  0.247&  0.128&  1.254\\
    3000 &  0.832&  0.554&  0.249&  0.128&  1.256\\
    5000 &  0.831&  0.553&  0.251&  0.129&  1.241\\
    7500 &  0.833&  0.553&  0.253&  0.131&  1.221\\
 10\,000 &  0.834&  0.554&  0.256&  0.132&  1.203\\
 15\,000 &  0.839&  0.557&  0.260&  0.134&  1.183\\
 20\,000 &  0.844&  0.561&  0.265&  0.136&  1.179\\
 30\,000 &  0.856&  0.569&  0.274&  0.141&  1.193\\
 50\,000 &  0.881&  0.585&  0.290&  0.149&  1.229\\
 75\,000 &  0.905&  0.601&  0.304&  0.155&  1.257\\
100\,000 &  0.919&  0.611&  0.312&  0.159&  1.270\\
\hline
  \multicolumn{1}{r}{T (K)}&
  \multicolumn{1}{c}{$2 - 4$}&
  \multicolumn{1}{c}{$2 - 5$}&
  \multicolumn{1}{c}{$3 - 4$}&
  \multicolumn{1}{c}{$3 - 5$}&
  \multicolumn{1}{c}{$4 - 5$}\\
\hline
     100 &  0.791&  0.315&  0.439&  0.308&  0.273\\
     150 &  0.793&  0.316&  0.440&  0.308&  0.274\\
     200 &  0.793&  0.316&  0.440&  0.309&  0.274\\
     300 &  0.794&  0.316&  0.440&  0.309&  0.274\\
     500 &  0.796&  0.317&  0.441&  0.310&  0.274\\
     750 &  0.797&  0.318&  0.442&  0.310&  0.275\\
    1000 &  0.799&  0.318&  0.443&  0.311&  0.275\\
    1500 &  0.801&  0.319&  0.444&  0.312&  0.276\\
    2000 &  0.804&  0.320&  0.445&  0.313&  0.276\\
    3000 &  0.809&  0.322&  0.448&  0.315&  0.277\\
    5000 &  0.820&  0.326&  0.454&  0.319&  0.279\\
    7500 &  0.834&  0.332&  0.462&  0.324&  0.282\\
 10\,000 &  0.851&  0.339&  0.472&  0.331&  0.285\\
 15\,000 &  0.891&  0.356&  0.494&  0.345&  0.294\\
 20\,000 &  0.930&  0.371&  0.516&  0.360&  0.305\\
 30\,000 &  0.997&  0.396&  0.551&  0.386&  0.327\\
 50\,000 &  1.084&  0.427&  0.595&  0.421&  0.361\\
 75\,000 &  1.144&  0.447&  0.624&  0.445&  0.388\\
100\,000 &  1.178&  0.458&  0.639&  0.459&  0.405\\
\hline
\end{tabular}
\end{flushleft}
\end{table}

As for the collision strengths $\Omega$, the ratio of
effective collision strengths
$\Upsilon$($^4$S$_{3/2}^{\mathrm o}$ -- $^2$D$_{5/2}^{\mathrm o}$) /
$\Upsilon$($^4$S$_{3/2}^{\mathrm o}$ -- $^2$D$_{3/2}^{\mathrm o}$)
for the transitions $1-2$ and $1-3$ remains constant, and
equal to about 1.5. Hence we do not detect any deviation from the ratio of
statistical weights of the upper levels, as found in the results of
\citet{BMM98}. This is also the case
for the ratio for the lines $1-4$ and $1-5$, which is very close to
2.0 and corresponds to the ratio of the statistical weights of
the upper levels $^2$P$_{3/2}^{\mathrm o}$ and $^2$P$_{1/2}^{\mathrm o}$.

\subsection{ 
\label{s:mesh}
Energy mesh for collision strengths
}

It is very important to use an energy mesh $\Delta E$ which will enable
us to delineate all important resonance structure. Since there are
resonances in the collision strengths very close to the excitation
thresholds (see Fig.~\ref{fig:omega12345}), a mesh which is not sufficiently
fine
could lead to
some inaccuracies when effective collision strengths
$\Upsilon$ are computed, especially at the lower end of the electron
temperature range.

\begin{table}
\begin{flushleft}
\caption{Comparison of effective collision strengths $\Upsilon$ for transitions
among the ground configuration 1s$^2$2s$^2$2p$^3$ levels of \mbox{O\,{\sc ii}},
calculated using
         the {\sl TFD} radial orbitals and the experimental target energies for
         various energy meshes;
           $h_1: \Delta E = 2 \times 10^{-5}$ Ry,
           $h_2: \Delta E = 1 \times 10^{-5}$ Ry,
           $h_3: \Delta E = 5 \times 10^{-6}$ Ry,
           $h_4: \Delta E = 2.5 \times 10^{-6}$ Ry.
}
  \label{tab:mesh.tfd}
  \begin{tabular}{rllll}
\hline
  \multicolumn{1}{r}{T (K)}&
  \multicolumn{1}{c}{$h_1$}&
  \multicolumn{1}{c}{$h_2$}&
  \multicolumn{1}{c}{$h_3$}&
  \multicolumn{1}{c}{$h_4$}\\
\hline
&  \multicolumn{4}{c}{$^4$S$_{3/2}^{\mathrm o} -^2$D$_{5/2}^{\mathrm o}$}\\
      100&  0.774&  0.786&  0.792&  0.796\\
      150&  0.783&  0.791&  0.795&  0.797\\
      200&  0.787&  0.794&  0.797&  0.798\\
      300&  0.793&  0.798&  0.800&  0.801\\
      500&  0.803&  0.806&  0.807&  0.808\\
     1000&  0.821&  0.822&  0.823&  0.823\\
     2000&  0.831&  0.831&  0.832&  0.832\\
  10\,000&  0.834&  0.834&  0.834&  0.834\\
 100\,000&  0.919&  0.919&  0.919&  0.919\\
\hline
&  \multicolumn{4}{c}{$^4$S$_{3/2}^{\mathrm o} - ^2$D$_{3/2}^{\mathrm o}$}\\
      100&  0.517&  0.525&  0.529&  0.531\\
      150&  0.523&  0.528&  0.531&  0.533\\
      200&  0.526&  0.530&  0.532&  0.533\\
      300&  0.530&  0.533&  0.534&  0.535\\
      500&  0.537&  0.538&  0.539&  0.540\\
     1000&  0.549&  0.549&  0.550&  0.550\\
     2000&  0.554&  0.554&  0.554&  0.555\\
  10\,000&  0.554&  0.554&  0.554&  0.554\\
 100\,000&  0.611&  0.611&  0.611&  0.611\\
\hline
&  \multicolumn{4}{c}{$^2$D$_{5/2}^{\mathrm o} - ^2$D$_{3/2}^{\mathrm o}$}\\
      100&  1.064&  1.082&  1.090&  1.095\\
      150&  1.065&  1.077&  1.083&  1.086\\
      200&  1.063&  1.072&  1.076&  1.078\\
      300&  1.061&  1.067&  1.070&  1.072\\
      500&  1.090&  1.094&  1.096&  1.097\\
     1000&  1.191&  1.193&  1.194&  1.194\\
     2000&  1.252&  1.253&  1.254&  1.254\\
  10\,000&  1.201&  1.205&  1.205&  1.203\\
 100\,000&  1.269&  1.271&  1.271&  1.270\\
\hline
\end{tabular}
\end{flushleft}
\end{table}

We have calculated collision strengths for different
values of the energy mesh $\Delta E$ in order to ensure the convergence
of our data. A comparison of effective collision strengths ($\Upsilon$)
is presented in Table~\ref{tab:mesh.tfd}. In this Table
we list data for transitions among
the three lowest fine-structure
levels of the configuration 1s$^2$2s$^2$2p$^3$, obtained by employing
four different values of energy mesh, the coarsest one being
$\Delta E = 2 \times 10^{-5}$ Ry and the finest one being
$\Delta E = 2.5 \times 10^{-6}$ Ry. Values for $\Upsilon$ are presented
for the transitions which have resonances lying close to the
first excitation threshold. For other
transitions, agreement is even better than for the ones presented here.

One can see from Table~\ref{tab:mesh.tfd} that the convergence
of effective collision strengths $\Upsilon$ with regard to
energy mesh is achieved. Any noticable difference in $\Upsilon$
does not exceed $4\%$ at the very low electron temperatures,
and it is negligible for temperatures above 1000~K.
Consequently, we are sure that the energy mesh $\Delta E$ applied
in our calculation is sufficiently fine to properly delineate the
resonance structure in the collision strengths, and that it does not lead to
any substantial inaccuracies in our computed data for effective collision
strengths.

\subsection{ 
\label{s:conf.set}
Comparison with other data
}

In addition to examining the influence of the energy mesh employed in
our calculation, we wish to examine how the choice of different
CI expansion for the target states and different radial orbital (RO) sets
affects the computed collision strengths and effective collision
strengths. Additionally,  we wish to investigate
how the use of the experimental target energies in the R-matrix calculations
can influence the results.

\begin{table*}
\begin{flushleft}
\caption{Comparison of the effective collision strengths $\Upsilon$ for transitions
         among the ground configuration 1s$^2$2s$^2$2p$^3$ levels in \mbox{O\,{\sc ii}}, 
         calculated using different configuration sets, different radial orbitals, 
         experimental and theoretical level energies, with the Breit-Pauli calculations 
         of \citet{mm2006} denoted BPRM06, \citet{BMM98} denoted as MB98, and 
	 \citet{tayal2007} denoted as T07.}
  \label{tab:compare}
  \begin{tabular}{rlllllllll}
\hline
 &\multicolumn{3}{c}{Experimental level energies}&
  \multicolumn{3}{c}{Theoretical level energies}&
  \multicolumn{3}{c}{Other calculation}\\
\cline{2-10}
  \multicolumn{1}{r}{T (K)}&
  \multicolumn{1}{l}{TFD}&
  \multicolumn{1}{l}{TFD1}&
  \multicolumn{1}{l}{STO1}&
  \multicolumn{1}{l}{TFD}&
  \multicolumn{1}{l}{TFD1}&
  \multicolumn{1}{l}{STO1}&
  \multicolumn{1}{c}{BPRM06}&
  \multicolumn{1}{c}{T07}&
  \multicolumn{1}{c}{MB98}\\
\hline
&  \multicolumn{9}{c}{$^4$S$_{3/2}^{\mathrm o} - ^2$D$_{5/2}^{\mathrm o}$}\\
     200& 0.798& 0.791& 0.778& 0.794& 0.812& 1.034&      &	&     \\
     500& 0.808& 0.802& 0.787& 0.797& 0.826& 0.961&      &	&     \\
    1000& 0.823& 0.818& 0.802& 0.805& 0.841& 0.909& 0.864&	&     \\
    5000& 0.831& 0.828& 0.812& 0.828& 0.837& 0.848& 0.885& 0.798& 0.81\\
 10\,000& 0.834& 0.830& 0.815& 0.835& 0.837& 0.846& 0.883& 0.803& 0.82\\
 20\,000& 0.844& 0.837& 0.823& 0.846& 0.843& 0.852& 0.885& 0.813& 0.84\\
100\,000& 0.919& 0.911& 0.898& 0.922& 0.908& 0.921&      & 0.874& 0.94\\
&  \multicolumn{9}{c}{$^4$S$_{3/2}^{\mathrm o} - ^2$D$_{3/2}^{\mathrm o}$}\\
     200& 0.533& 0.529& 0.519& 0.529& 0.541& 0.705&      &	&     \\
     500& 0.540& 0.536& 0.526& 0.532& 0.551& 0.656&      &	&     \\
    1000& 0.550& 0.547& 0.536& 0.538& 0.563& 0.616& 0.590&	&     \\
    5000& 0.553& 0.550& 0.539& 0.553& 0.558& 0.568& 0.587& 0.548& 0.41\\
 10\,000& 0.554& 0.551& 0.541& 0.559& 0.560& 0.567& 0.585& 0.550& 0.43\\
 20\,000& 0.561& 0.557& 0.547& 0.566& 0.564& 0.571& 0.585& 0.553& 0.44\\
100\,000& 0.611& 0.605& 0.597& 0.614& 0.604& 0.613&      & 0.585& 0.49\\
&  \multicolumn{9}{c}{$^4$S$_{3/2}^{\mathrm o} - ^2$P$_{3/2}^{\mathrm o}$}\\
     200& 0.245& 0.245& 0.251& 0.248& 0.258& 0.250&      &	&     \\
     500& 0.245& 0.246& 0.252& 0.249& 0.258& 0.250&      &	&     \\
    1000& 0.246& 0.246& 0.252& 0.249& 0.259& 0.251& 0.299&	&     \\
    5000& 0.251& 0.251& 0.257& 0.254& 0.263& 0.255& 0.307& 0.279& 0.25\\
 10\,000& 0.256& 0.256& 0.261& 0.259& 0.267& 0.260& 0.313& 0.283& 0.26\\
 20\,000& 0.265& 0.264& 0.269& 0.268& 0.274& 0.268& 0.322& 0.288& 0.27\\
100\,000& 0.312& 0.309& 0.310& 0.314& 0.313& 0.313&      & 0.315& 0.33\\
&  \multicolumn{9}{c}{$^4$S$_{3/2}^{\mathrm o} - ^2$P$_{1/2}^{\mathrm o}$}\\
     200& 0.126& 0.126& 0.131& 0.128& 0.129& 0.134&      &	&     \\
     500& 0.127& 0.127& 0.131& 0.128& 0.129& 0.134&      &	&     \\
    1000& 0.127& 0.127& 0.132& 0.129& 0.129& 0.135& 0.148&	&     \\
    5000& 0.129& 0.129& 0.134& 0.131& 0.131& 0.137& 0.151& 0.138& 0.11\\
 10\,000& 0.132& 0.132& 0.136& 0.133& 0.134& 0.139& 0.152& 0.140& 0.12\\
 20\,000& 0.136& 0.136& 0.139& 0.138& 0.138& 0.142& 0.156& 0.142& 0.12\\
100\,000& 0.159& 0.158& 0.159& 0.161& 0.160& 0.160&      & 0.157& 0.15\\
&  \multicolumn{9}{c}{$^2$D$_{5/2}^{\mathrm o} - ^2$D$_{3/2}^{\mathrm o}$}\\
     200& 1.078& 1.145& 1.261& 1.289& 1.035& 3.560&      &      &     \\
     500& 1.097& 1.145& 1.218& 1.709& 1.138& 2.574&      &      &     \\
    1000& 1.194& 1.231& 1.270& 1.699& 1.275& 1.957& 1.618&      &     \\
    5000& 1.241& 1.258& 1.255& 1.404& 1.273& 1.349& 1.518& 1.653& 1.52\\
 10\,000& 1.203& 1.211& 1.202& 1.298& 1.218& 1.250& 1.426& 1.434& 1.25\\
 20\,000& 1.179& 1.176& 1.158& 1.234& 1.179& 1.200& 1.324& 1.291& 1.17\\
100\,000& 1.270& 1.260& 1.241& 1.281& 1.255& 1.274&      & 1.260& 1.24\\
\hline
\end{tabular}
\end{flushleft}
\end{table*}

In Table~\ref{tab:compare} we present the effective collision strengths
$\Upsilon$ obtained using different configuration and target energy
sets, and compare our results with available data from other authors.
One set of data is obtained using adjusted (to the experimental)
target energies.
The target level energy corrections $\Delta E_{\mathrm{TFD}}$,
$\Delta E_{\mathrm{TFD1}}$ and $\Delta E_{\mathrm{STO1}}$ in this type
of calculation for the different sets of radial orbitals are presented in
Table~\ref{tab:target}. In another set of calculations, the pure
theoretical results of the {\it ab initio} energy levels are used,
without any adjustment being applied to the target level energies.

For each set of target level energies, we have performed three series of
calculations. In the calculation denoted {\sl TFD} we use the most extensive
set of configurations in the target wavefunction CI expansion, which
is given in Sec.~\ref{s:target}, based on Thomas-Fermi-Dirac type radial orbitals.
{\sl TFD1} uses a smaller wavefunction CI expansion
(see Sec.~\ref{s:target}) with the same radial TFD-type orbitals. The third
calculation {\sl STO1} uses the same configuration set as
{\sl TFD1}, but employs Slater-type radial orbitals as
a basis. This type of calculation is the closest one to that performed
by \citet{BMM98}.

We compare our data for $\Upsilon$ with the results of \citet{mm2006} 
which we denote as {\sl BPRM06}. Their close-coupling calculation was performed 
in the Breit-Pauli approximation using the R-matrix codes. The results given
by \citet{tayal2007} obtained by using  a 47-level Breit-Pauli R-matrix approach
with nonorthogonal radial functions are presented in the column {\sl T07}.
In addition, we include data from \citet{BMM98}, denoted as {\sl MB98}, 
for comparison in  the last column of Table~\ref{tab:compare}. 
There are two main differences between our calculation and that of 
\citet{mm2006}: 
(i) They used a smaller CI expansion of the target, 6 configurations compared 
to our 22;
(ii) Their 3p and 3d radial functions were real physical orbitals, whereas ours 
are correlation orbitals optimized to improve the representation of the 
2s$^2$2p$^3$ and 2s2p$^4$ levels.

When comparing our results obtained using the experimental target energies
but different sets of configuration expansion and different radial orbitals,
we can see that the values for $\Upsilon$ show no substantial differences.
The {\sl TFD} and {\sl TFD1} data almost exactly match, while
the {\sl STO1} results differ by a few percent at lower
temperatures for the transitions originating from the ground level
$^4$S$_{3/2}^{\mathrm o}$. There are minor differencies in the effective
collision strengths for the
$^2$D$_{5/2}^{\mathrm o} - ^2$D$_{3/2}^{\mathrm o}$
transition at very low electron temperatures, but this becomes
negligible for $T \geq 5000$~K.

For the results obtained using the {\it ab initio} target levels energies,
the situation is quite different. There are noticeable discrepancies
in the values of $\Upsilon$ for the different configuration expansion sets
and different radial orbitals. This is particularly true for
$^4$S$_{3/2}^{\mathrm o} - ^2$D$_{5/2}^{\mathrm o}$,
$^4$S$_{3/2}^{\mathrm o} - ^2$D$_{3/2}^{\mathrm o}$ and
$^2$D$_{5/2}^{\mathrm o} - ^2$D$_{3/2}^{\mathrm o}$, and for
low electron temperatures. There is much better
agreement at higher temperatures for these transitions
as well as for
$^4$S$_{3/2}^{\mathrm o} - ^2$P$_{3/2}^{\mathrm o}$ and
$^4$S$_{3/2}^{\mathrm o} - ^2$P$_{1/2}^{\mathrm o}$,
where the discrepancies in $\Upsilon$ are very small at all temperatures.
Results obtained using the theoretical target level
energies are generally consistent, and agree
with the
data obtained using the experimental target energies.
Difference are due to the resonances positioned very close
to the excitation threshold. Their position depends on the type of radial orbitals
used and on the CI expansion applied (see Table~\ref{tab:target}).

\begin{figure}
\resizebox{1.0\hsize}{!}{\includegraphics{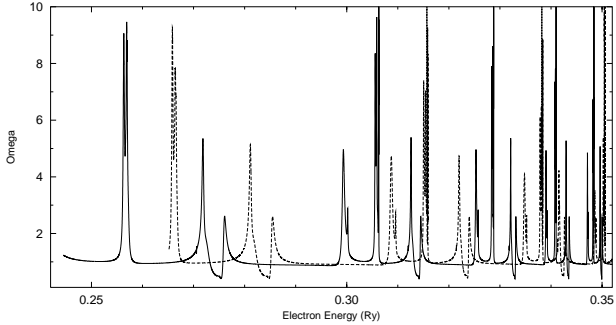}}
\caption{Collision strengths for the transition
$^2$D$_{5/2}^{\mathrm o}$ -- $^2$D$_{3/2}^{\mathrm o}$
in \mbox{O\,{\sc ii}} near the excitation threshold.
Electron energies are in Rydbergs relative to the
ground level $^4$S$_{3/2}^{\mathrm o}$. The solid line is for
the calculation with adjusted target level energies, while the
dashed
line is for data with {\em ab initio} level energies.
\label{fig:omega23}}
\end{figure}

For illustrative purposes, we present a plot of the near-threshold
collision strengths $\Omega$ for
$^2$D$_{5/2}^{\mathrm o}$ -- $^2$D$_{3/2}^{\mathrm o}$
obtained within the {\sl STO1} set, using both the experimental (solid line)
and the theoretical (dashed line) target level energies in Fig.~\ref{fig:omega23}.
It is clear that the first resonance structure is located right on
the edge of the excitation threshold, at 0.26519~Ry
(see Table~\ref{tab:target}) in the case of the theoretical target energies.
However, the same resonance structure is shifted away from the
threshold by more than 0.01~Ry when energy adjustments are introduced.
Even if the background value of collision strength does not depend on the type
of target energies used (as may be seen from Fig.~\ref{fig:omega23}), 
the low electron temperature behaviour of the effective collision strength 
$\Upsilon$ is defined by the near-threshold resonances and their positions.

When the electron temperature increases, the low-energy part of the collision
strength $\Omega$ becomes less important in the overall value of $\Upsilon$,
and the agreement of the different sets of effective collision strengths
$\Upsilon$  becomes significantly better. This points to the fact that
collision strength background values are essentially the same both
for the experimental and theoretical target level energies.
Hence, introducing the target-energy adjustments changes only the
positions of resonances. Therefore, these adjustments cannot lead to
substantial deviations for the calculated effective collision strengths,
especially at the higher electron temperatures.

A comparison with the Breit-Pauli results ({\sl BPRM06}) of \citet{mm2006} and 
\citet{ap2005} indicates reasonable agreement, although our $\Upsilon$ values 
are consistently smaller than their values. For the transition 
$^4$S$_{3/2}^{\mathrm o} - ^2$D$_{5/2}^{\mathrm o}$, the difference is $4-5\%$, 
for $^4$S$_{3/2}^{\mathrm o} - ^2$D$_{3/2}^{\mathrm o}$ is $5-8\%$, 
for $^4$S$_{3/2}^{\mathrm o} - ^2$P$_{3/2}^{\mathrm o}$ is around $20\%$ 
and for $^4$S$_{3/2}^{\mathrm o} - ^2$P$_{1/2}^{\mathrm o}$ is $15-17\%$. These
discrepancies are caused by the different background values of the corresponding
collision strengths $\Omega$, arising from the different configuration expansion
sets used in our calculation and in those of \cite{mm2006}.

A slightly more complicated situation is observed for
$^2$D$_{5/2}^{\mathrm o} - ^2$D$_{3/2}^{\mathrm o}$, where the
discrepancy in $\Upsilon$ is $36\%$ at $T = 1000$~K, which falls to
just $12\%$ at $T = 20\,000$~K. We can attribute the larger
discrepancy of the low temperature results to differences in the
resonance structure of the collision strengths positioned right on the
excitation threshold of this transition, which is noticeable in fig.~1
from \citet{ap2005}.  The discrepancies at higher
temperatures are for the same reason as for the transitions
originating from the ground level 2s$^2$2p$^3$ $^4$S$_{3/2}^{\mathrm
o}$.

A similar pattern can be observed for transitions originating from
the levels $^2$D$_{5/2}^{\mathrm o}$ and $^2$D$_{3/2}^{\mathrm o}$,
where differences in the calculated values of $\Upsilon$
remain approximately
constant, and are largely due to the differing background values.
When we compare effective collision strength ratios
for the transitions originating from the ground level $^4$S$_{3/2}^{\mathrm o}$,
we see that they are very close to the ratios of the statistical weights
of the upper levels. At $T = 10000$~K, the ratio
$\Upsilon$($^4$S$_{3/2}^{\mathrm o} - ^2$D$_{5/2}^{\mathrm o}$)/
$\Upsilon$($^4$S$_{3/2}^{\mathrm o} - ^2$D$_{3/2}^{\mathrm o}$)
is 1.505 in our calculation and 1.509 in that of \citet{ap2005}, while the ratio
$\Upsilon$($^4$S$_{3/2}^{\mathrm o} - ^2$P$_{3/2}^{\mathrm o}$)/
$\Upsilon$($^4$S$_{3/2}^{\mathrm o} - ^2$P$_{1/2}^{\mathrm o}$)
is 1.94 and 2.06, respectively. Although for transitions
originating from $^2$D$_{5/2}^{\mathrm o}$ and
$^2$D$_{3/2}^{\mathrm o}$ to $^2$P$_{3/2}^{\mathrm o}$ and
$^2$P$_{1/2}^{\mathrm o}$, the ratio of $\Upsilon$ does not
correspond to the ratio of the upper level statistical weights,
it is approximately the same both in our calculation and in that of
{\sl BPRM06}.

A comparison with data of \citet{tayal2007} presented in the column {\sl T07} 
of Table~\ref{tab:compare} indicates very good agreement. For the transitions 
$^4$S$_{3/2}^{\mathrm o} - ^2$D$_{5/2}^{\mathrm o}$, $^2$D$_{3/2}^{\mathrm o}$, 
the differences in $\Upsilon$ values do not exceed few percent, with our results
being slightly higher. For the transitions to the levels 
$^2$P$_{3/2}^{\mathrm o}$ and $^2$P$_{1/2}^{\mathrm o}$ our calculated effective 
collision strengths are slightly smaller than those of \citet{tayal2007}. 
A similar pattern is observed for the forbidden transitions not only from 
the ground level but also from the excited levels of the multiplets $^2$D and 
$^2$P. However, we note that the results for the excitation to the levels 
$^2$P$_{3/2}^{\mathrm o}$ and $^2$P$_{1/2}^{\mathrm o}$ in table 3 of 
\citet{tayal2007} should be swapped to obtain the correct data. 
This was probably due to the fact that the energy ordering for the corresponding 
calculated levels differs from the experimental one.
The effective collision strengths for the transition 
$^2$D$_{5/2}^{\mathrm o} - ^2$D$_{3/2}^{\mathrm o}$ differ significantly, 
the deviation reaching some $25\%$ at the lower temperature end. 
It should be noted that the differences are smaller when the theoretical
level energies are employed in our calculations, suggesting that the main 
reason for this disagreement is that we use experimental level energies
in our scattering calculation leading to more accurate data. 
This is particularly important for the transition 
$^2$D$_{5/2}^{\mathrm o} - ^2$D$_{3/2}^{\mathrm o}$
because of resonance structures present very close to the excitation threshold
(see Fig.~\ref{fig:omega23}).

A comparison of our data (from the {\sl TFD} set) with the Breit-Pauli calculations 
of \citet{BMM98} reveals two different trends. For some transitions, namely
$^4$S$_{3/2}^{\mathrm o} - ^2$D$_{5/2}^{\mathrm o}$ and
$^4$S$_{3/2}^{\mathrm o} - ^2$P$_{3/2}^{\mathrm o}$,
the agreement is exceptionally good, even better than the \citet{mm2006} data. 
This is due to the similar target and wavefunction CI expansion used.
For $^2$D$_{5/2}^{\mathrm o} - ^2$D$_{3/2}^{\mathrm o}$, the effective collision 
strengths agree very well at higher electron temperatures, whereas some
difference appears at $T = 5000$~K, which can be attributed to
the effect of the near-threshold resonances.

For other transitions from the ground state, shown in
Table~\ref{tab:compare}, there are very significant discrepancies
between our effective collision strengths and those of \citet{BMM98}.
The data differ by nearly 30\% for
excitation to the $^2$D$_{3/2}^{\mathrm o}$ level and by
15--20\% for the $^2$P$_{1/2}^{\mathrm o}$ level.
It is worth noting that both of these are the upper levels of their
corresponding terms, $^2$D or $^2$P, respectively.
Such a large drop in the value of the effective collision strength
$\Upsilon$ causes the significant deviation from the statistical
weights ratio, which is not expected for a singly-ionized
ion with $Z = 8$. McLaughlin \& Bell explain
this effect by the influence of configuration mixing, but there
are no data presented in their work to confirm such a conclusion.

Checking our CI wavefunction expansion coefficients for the
fine-structure levels of the 1s$^2$2s$^2$2p$^3$ configuration,
we do not find any substantial configuration mixing effects which can cause
this kind of deviation. The main contributing term is usually more than 0.95
for levels with $J=3/2$ and approximately 0.99 for levels
with $J=1/2$. These appear to be very reasonable values for a
low-Z ion. Therefore, we conclude that the data of \citet{BMM98} for some 
transitions are incorrect. Although we cannot define any particular reason 
for the inaccuracy of their data, the most plausible cause is a limited
CI expansion of the target wavefunctions, where an incomplete set of CSFs 
is used.

\section{
Conclusions
}

\begin{figure}
\resizebox{1.0\hsize}{!}{\includegraphics{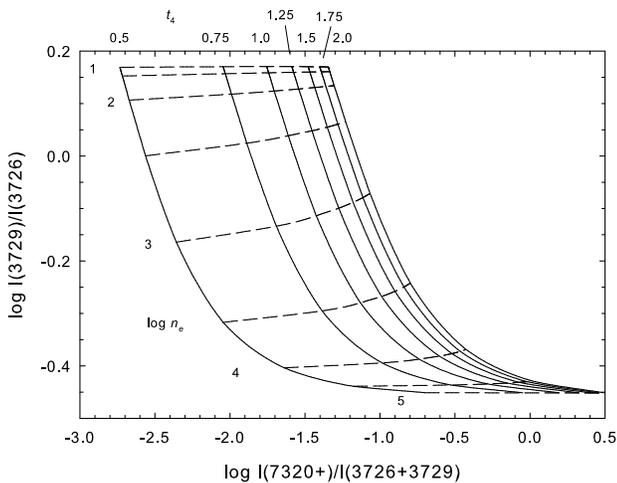}}
\caption{Ratio -- ratio diagram for \mbox{O\,{\sc ii}} transitions,
where I is in energy units, computed using 
the new collision strengths
and the transition probabilities described in the text.
The electron density $n_e$ is in units of cm$^{-3}$,
while the electron temperature $t_4$ is in units of 10$^{4}$\,K.
\label{fig:ratios} }
\end{figure}

In the current work we have determined the collision strengths $\Omega$
and the effective collisions strengths $\Upsilon$ for a wide range
of electron temperatures $T$ using the relativistic Breit-Pauli R-matrix
code for the excitation of forbidden lines among the fine-structure
levels of the ground configuration 1s$^2$2s$^2$2p$^3$ of the O$^+$ ion.
The collision strengths are calculated using a very fine energy mesh,
which allows the delineation of all resonance structure to high accuracy.
A comparison of the effective collision strengths obtained using
different energy meshes confirms that a convergence of $\Omega$ on the energy 
mesh was achieved.

The collision strengths are computed using various target
wavefunction expansions and different sets of the radial orbitals,
employing both {\em ab initio} theoretical and the experimental energies
for the target levels. In all cases we do not detect any sizeable
difference in the  background value of the calculated $\Omega$.

In all six datasets for our calculations, we do not find any
significant departure from the statistical distribution for the
ratio of the collision strengths $\Omega$ and the effective collision
strengths $\Upsilon$. This confirms the findings of \citet{mm2006}, 
and shows that the results of \citet{BMM98} for some transitions are inaccurate. 
Although we have tried to replicate the latter calculation and establish 
the origin of the departure of their results from the statistical weight rule, 
we did not find any reason why it could happen.

Consequently, any analysis of observations based on the atomic data 
from \citet{BMM98} must be treated with caution. The differences between our 
results and those of \citet{mm2006} and \cite{ap2005} can be attributed to 
the more extensive and converged CI expansion used here and we
therefore consider the results given in Table~\ref{tab:upstfd} to be the best
available for this ion at present.

The revised rates will change the plasma diagnostics presented by \citet{fpk1999}.
We have threfore regenerated the \mbox{O\,{\sc ii}} line ratios and show some results 
in Fig.~\ref{fig:ratios}. These employ the transition probabilities given 
by \citet{zeippen82} since these are in better agreement with observations 
\citep{wangliu}. We refer the reader to \citet{fpk1999} for further details, 
and to compare how the new collision rates have changed the results.

\section*{
Acknowledgments
}

Support for proposal HST-AR-09923 was provided by NASA through a grant from
the Space Telescope Science Institute, which is operated by the Association
of Universities for Research in Astronomy, Inc., under NASA contract
NAS5-26555.
FPK is grateful to AWE Aldermaston for the award of a William Penney Fellowship.
This work was supported by STFC
and EPSRC, and also by NATO Collaborative
Linkage Grant CLG.979443. We are also grateful to the Defence Science and Technology
Laboratory (dstl) for support under the Joint Grants
Scheme.
GJF thanks the NSF (AST 0607028), NASA (NNG05GD81G), STScI
(HST-AR-10653) and the Spitzer Science Center (20343) for support.
We acknowledge the use of software from the Condor Project
(http://www.condorproject.org/) in running the R-matrix codes.

\label{lastpage}

\end{document}